\documentclass[prd,twocolumn,floatfix,preprintnumbers,letterpaper]{revtex4}
\usepackage{graphicx}
\usepackage{amsmath,amssymb}
\input{epsf}
\usepackage{epsf}

\begin{document}
\title{Cosmological Scaling Solutions with Tachyon: Modified Gravity model}
\author{A.A. Sen {\footnote{anjan.ctp@jmi.ac.in}}and N. Chandrachani Devi\footnote{chandrachani@gmail.com}}
\affiliation{Center For Theoretical Physics, Jamia Milia Islamia,
New Delhi 110025, India}

\begin{abstract}
Modifying the Einstein's gravity at large distance scales is one of the interesting proposals to explain the late time acceleration of the universe. In this paper, we analyse  scaling solutions in modified gravity models where the universe is sourced by a background matter fluid together with a tachyon type scalar field. We describe a general prescription to calculate the scaling potential in such models. Later on we consider specific examples of modifications and apply our method to calculate the scaling potential and the scale factor. Our method can be applied to any modified gravity model, in  presence of tachyon field.
\end{abstract}
\maketitle
\vspace{5mm}

\section{Introduction}
Accelerated expansion of the universe has always been one of the most challenging subjects of investigation in cosmology. For early universe, it plays an important role for solving major problems in standard cosmology like horizon and flatness problem. It also provides an effective mechanism for producing a nearly scale-invariant spectra for quantum fluctuations of the inflaton  which can work as a seed for the structure formations in our universe. This paradigm which is dubbed as ``Inflation'' plays a major role in modern era of cosmology \cite{infl}. On the other hand, there are not many daunting questions in cosmology today  than what is the driving force behind the late time acceleration of the universe \cite{accl}. First directly observed by Supernova type Ia measurements by two independent groups in 1998 \cite{snold}, today it has also been confirmed by observations from new  SupernovaType Ia measurements\cite{snnew}, Cosmic Microwave Background Radiations (CMBR)\cite{cmbr}, and galaxy clustering by large scale redshift surveys\cite{lss}.

To discover the nature of driving force behind this late time acceleration of the universe, also termed in literature as ``dark energy'', is one of the most important tasks in cosmology today. The first and the most simple choice for dark energy is the vacuum energy or cosmological constant $\Lambda$. However this
possibility of {\bf $\Lambda$} being  dominant component of the total
energy density of the universe, has problem that the energy scale
involved is lower than normal energy scale of most particle
physics models by a factor of $\sim 10^{-123}$. 

An optimistic choice for modelling the missing energy of the universe is in the form of a scalar field with a canonical kinetic energy, slowly rolling down a considerably flat potential so that its slowly varying energy density mimics an effective cosmological constant. This form of missing energy density is  popularly called `` Quintessence''\cite{quint}. It is similar to the inflaton field with the difference that it evolves in a much lower energy scale. Amongst them, tracker quintessence models have the advantage of allowing the late time accelerating epoch to be reached from a wide range of initial conditions. In order to have a viable dark energy model, it is important that this scalar field remains subdominant  during most of the thermal history of the universe. In this sense, it is necessary that this scalar field mimicks the background energy density in the early time. This property is called the ``scaling or attractor  property''. This property is very important in cosmology as they allow us to study the asymptotic behavior of a particular cosmology and check whether that particular behavior is stable or not. There have been numerous investigations involving scaling solutions in cosmology in standard Einstein's gravity as well as in scalar tensor gravity \cite{scaling}.

Recently, there are suggestions that the present acceleration of the universe is not due to any new unknown component in the cosmic soup, but due to the modifications of the gravitational physics at large distance scales. In this regard, possible modifications in the Friedmann's equation have  been proposed. Most of these modifications have been inspired by the higher dimensional brane-world models. A few suggestions are particularly interesting in this regard. The Dvali-Gabadadze-Poratti (DGP) brane induced gravity model\cite{dgp}, the Cardassian model proposed by Freese and Lewis\cite{card}, a model by Dvali and Turner\cite{turner}, by Shtanov and Sahni\cite{shtanov} and also the modified Chaplygin gas model by Barriero and Sen\cite{mgcg} are some of them. Cosmological solutions with a canonical scalar field have been studied in the modified gravity models\cite{scalemod}. 

Recently, an effective scalar field theory governed by a Lagrangian density with a non-canonical kinetic energy term ${\cal{L}} = -V(\phi)F(X)$, where $X = -(1/2)\partial^{\mu}\phi\partial_{\mu}\phi$, has attracted considerable attention in cosmology. Such model can lead to a late-time acceleration and is called ``k-essence''\cite{kessen}. A scalar field with a non-canonical kinetic term has also been investigated for an early universe inflationary scenario and is termed as ``k-inflation''\cite{kinf}. One example of such Lagrangian density is ${\cal L}_{tach}=-V(\phi)\sqrt{1- \partial^{\mu}\phi\partial_{\mu}\phi}$. As discussed by Padmanabhan and Roy Choudhury\cite{paddytirth}, this is generalization of the Lagrangian of a relativistic particle. The Hamiltonian structure of tachyonic matter given by the above Lagrangian density is very similar to that of a special relativistic particle governed by ${\cal L} = -m\sqrt{1-\dot{q}^2}$ where $m$ and $q$ are the mass and generalized coordinate of the particle respectively. This tachyon field can naturally arise in open string theory\cite{asoke} and can provide a rich gamut of possibilities in cosmological context\cite{harry}. Padmanabhan and choudhury \cite{paddytirth} also explored the possiblity of this kind field acting both as a clustered dark matter and smooth dark energy with a scale dependent equation of state (see \cite{mota} for a different approach for such possibility).

In this paper we have studied systematically the scaling solutions involing a tachyon field described by the Lagrangian density,mentioned above, in modified version of Friedman equation. In section 2, we describe the equations of motion in the modified version of gravity in general and define the variables that allow us to study the scaling solutions. The conditions for scaling behavior are introduced in Section 3 and we also describe how to calculate the potentials for the tachyon field which describes the scaling behavior. In section 4, we apply our method to some particular examples of the modified gravity e.g the Cardassian model, the Shtanov-Sahni model, the Randall-Sundrum model and the DGP model. We conclude in section 5.

\vspace{8mm}
\section{Equation Of Motion In Modified Gravity}

We consider a spatially flat Friedmann-Robertson-Walker (FRW) model. In this case the expansion rate of the universe $H$ (also called the ``Hubble parameter'') is defined as
\begin{equation}
H^{2} = ({8\pi G/3})\rho L^{2}(\rho),
\end{equation}
\vspace{2mm}

\noindent
where $H \equiv {\dot{a}\over{a}}$ is the Hubble parameter, $a$ is the scale factor, $\rho$ is the total energy density of the universe. A ``dot'' denotes derivative with respect to 
cosmic time and $G$ is the gravitational constant. Whatever modification to standard Einstein's gravity we assume, is parametrized by the correction term $L(\rho)$ and this is assumed to be positive definite without any loss of generality. For $L(\rho)=1$, one recovers the standard Einstein's gravity.

Our aim is to investigate models where the universe is sourced by the matter component together with a dark energy part. In this paper, we assume that this dark energy field has a non-canonical kinetic energyof Dirac-Born-Infeld form with a Lagrangian density
\begin{equation}
{\cal L}_{tach}=-V(\phi)\sqrt{1- \partial^{\mu}\phi\partial_{\mu}\phi},
\end{equation}
\vspace{2mm}

\noindent
where $V(\phi)$ is the potential for the field $\phi$. To start with, we assume the matter part is described by a barotropic fluid with a equation of state $P_{\gamma} = (\gamma-1)\rho_{\gamma}$, $\gamma$ being a constant. Then the total energy density of the universe $\rho$, appearing the equation (1) is given by $\rho = \rho_{\gamma} + \rho_{\phi}$. One can calculate the energy density and pressure for the tachyon field $\phi$ from the Lagrangian density given in equation (2) as
\begin{eqnarray}
\rho_{\phi} &=& {V(\phi)\over{\sqrt{1-\dot{\phi}^2}}}\nonumber\\
P_{\phi} &=& -V(\phi)\sqrt{1-\dot{\phi}^2}.
\end{eqnarray}
\vspace{2mm}

\noindent
We also assume that as in standard gravity, the energy momentum tensors of matter and dark energy are covariantly conserved separately which implies that
\begin{eqnarray}
\dot{\rho_{\gamma}} &=& - 3H\gamma\rho_{\gamma}\nonumber\\
\ddot{\phi}&+& 3H\dot{\phi}(1-\dot{\phi}^2)+{V^{'}\over{V}}(1-\dot{\phi}^{2}) = 0.
\end{eqnarray}
\vspace{2mm}

\noindent
Equations (1) and (4) close the system of equations. We now define three new variables:
\begin{eqnarray}
X &=& \dot{\phi}\nonumber\\
Y &=& \sqrt{V\over{\rho}}\\
N &=& Log[a]\nonumber.
\end{eqnarray}
\vspace{2mm}

\noindent
When we expressed in terms of these new variables, the system of equations (1) and (4) becomes
\begin{eqnarray}
X^{'} &=& -(1-X^2)[3X-\sqrt{3}Y\lambda]\nonumber\\
Y^{'} &=& {Y\over{2}}\left[-\sqrt{3}\lambda XY - {3Y^{2}(\gamma-X^{2})\over\sqrt{1-X^{2}}}
 + 3\gamma\right]\\
\lambda^{'} &=& - \sqrt{3}\lambda^{2}XY(\Gamma - 3/2)\nonumber \\ 
&+&3\lambda\left[\gamma - {Y^{2}(\gamma-X^2)\over{\sqrt{1-X^{2}}}}\right]\rho {d Log[L(\rho)]\over{d\rho}}\nonumber,
\end{eqnarray}
\vspace{2mm}

\noindent
where ``prime'' means differentiation with respect to $N$. The parameter $\lambda$ and $\Gamma$ are defined as
\begin{eqnarray}
\lambda &=& -{(dV/d\phi)\over{\sqrt{8\pi G}L(\rho)V^{3/2}}}\nonumber\\
\Gamma &=&  {Vd^{2}V/d\phi^2\over{(dV/d\phi)^2}}.
\end{eqnarray}
\vspace{2mm}

\noindent
From the definition of total energy density, one can get the constraint equation
\begin{equation}
{Y^{2}\over{\sqrt{1-X^{2}}}} + {\rho_{\gamma}\over{\rho}} = 1.
\end{equation}

The equation of state for the tachyon field and its density parameter are given by
\begin{eqnarray}
\gamma_{\phi} &=& X^{2}\nonumber\\
\Omega_{\phi} &=& {Y^{2}\over{\sqrt{1-X^{2}}}}.
\end{eqnarray}
\vspace{2mm}

\noindent
Then the allowed phase space for $X$ and $Y$ is given by $0\leq X^{2}+Y^{4}\leq 1$ from the requirement $0\leq \Omega_{\phi} \leq 1$. From the expression of the equation of state $\gamma_{\phi}$, one see that $\gamma_{\phi} \geq 0$ ($P_{\phi} = (\gamma_{\phi} -1)\rho_{\phi}$).

For the case $\lambda = constant$, in standard gravity, the potential turns out to be $V(\phi) \propto \phi^{-2}$. But due to the presence of the modified term $L(\rho) \neq 1$, it is not so in our case. But it is interesting to see that, for $\lambda =constant$, the equations in (6) have the identical form to that of the plane autonomous system of standard relativistic cosmology involving  a tachyon field in terms of the parameters
\begin{eqnarray}
X_{s} &=& \dot{\phi}\nonumber\\
Y_{s} &=& {\sqrt{8\pi G V(\phi)}\over{\sqrt{3}H}}\\
\lambda_{s} &=& -{(dV/d\phi)\over{\sqrt{8\pi G}V^{3/2}}}\nonumber.
\end{eqnarray}
\vspace{2mm}

\noindent
This immediately ensures that our system of equations (6) with constant $\lambda$ admits the same set of critical points to that of the standard cosmology when those solution are expressed in terms of $\{X,Y,\lambda\}$. Also the stability of these critical points can be directly obtained from stability analysis of the standard cosmology scenario. Now there are altogether five fixed points in our system equations (6) when $\{X,Y,\lambda\} = \{X_{c}, Y_{c}, \lambda_{c}\}$ are constants. Three of these $\{X_{c}=0,Y_{c}=0\}$, $\{X_{c}=1,Y_{c}=0\}$ and $\{X_{c}=-1,Y_{c}=0\}$ are unstable as studied by Copeland et al\cite{copeland1}. The fourth fixed point is given by
\begin{eqnarray}
X_{c} &=& \sqrt{\gamma},\nonumber\\
Y_{c} &=& \sqrt{3 \gamma}/\lambda_{c}.
\end{eqnarray} 
\vspace{2mm}

\noindent
It is a stable one for $0\leq \gamma \leq \alpha = {\lambda_{c}^{2}\over{18}}\left(\sqrt{\lambda_{c}^{4}+36}-\lambda_{c}^{2}\right)$. For this one can easily show that $\gamma_{\phi} = \gamma$, hence it represents a scaling solution. But in this case as $\Omega_{\phi} = 3\gamma/(\lambda_{c}^{2}\sqrt{1-\gamma})$, $\gamma$ has to be less than 1. It restricts this case to be a viable model for dark energy as the background fluid can never be matter-like. The fifth fixed point is given by
\begin{eqnarray}
X_{c} &=& {\lambda_{c}\over{\sqrt{3}}} \left({{\sqrt{\lambda_{c}^{4}+36}-\lambda_{c}^{2}}\over{6}}\right)^{1/2}\nonumber\\
Y_{c} &=& \left({{\sqrt{\lambda_{c}^{4}+36}-\lambda_{c}^{2}}\over{6}}\right)^{1/2}
\end{eqnarray}
\vspace{2mm}

\noindent
This fixed point is a stable node for $\gamma \geq \alpha\equiv {\lambda_{c}^{2}\over{18}}\left(\sqrt{\lambda_{c}^{4}+36} -\lambda_{c}^{2}\right)$.
\vspace{5mm}

\section{Scaling Solutions}

As we mentioned in the earlier section, that for a standard cosmology, $\lambda = const$, corresponds to a potential of inverse power law form. But with the modification term $L(\phi)$ in the expression of $\lambda$ in equation (6), the form of the potential will change. In this section, we shall calculate the form of the scaling potential in the modified gravity scenario. For this purpose, we essentially follow the method earlier prescribed by Copeland {\it et al.}\cite{copeland2} for a standard canonical scalar field.

One can show that for both the critical points (11) and (12), $X^{'} = Y^{'} = \lambda^{'} = 0$, if the condition
\begin{equation}
\Gamma = {3\over{2}} + \rho{d\over{d\rho}}Log[L(\rho)]
\end{equation}
\vspace{2mm}

\noindent
is satisfied. Using the relation $\rho = V/Y_{c}^2$, the above equation can be written as
\begin{equation}
{d\over{d\phi}}(Log[{d\rho\over{d\phi}}]) - {3\over{2}}{d\over{d\phi}}\left[Log(\rho)\right] -{d\over{d\phi}}\left(Log[L(\rho)]\right) = 0.
\end{equation}

This is very similar to the corresponding equation (eqn (19) in ref \cite{copeland2}) for a canonical scalar field except the factor (3/2) in the second term in left hand side of the above equation. The above equation can be integrated to find $\rho(\phi)$ which together with the expression $V = {Y_{c}^2}\rho$ will give the corresponding potential for the scaling solution for given choice of modification $L(\rho)$. By integrating  the above equation and using equation (7) with constant $\lambda$, we get
\begin{equation}
\int {d\rho\over{L(\rho)\rho^{3/2}}} = - \kappa \lambda_{c} Y_{c}  \phi,
\end{equation}
\vspace{2mm}

\noindent
where $\kappa^{2} = 8\pi G$. While deriving the above expression, we have put one integration constant to zero without no loss of generality. This is equivalent to giving a linear shift to the value of the scalar field.

It is also interesting to calculate the behavior of the scale factor for a given scaling solution assuming the scalar field is monotonically varying function of time $(\dot{\phi} \neq 1)$. In general one can express the scalar field equation (4) in the following manner:
\begin{equation}
\dot{\rho_{\phi}} = {V(\phi)\over{\sqrt{1-\dot{\phi}^2}}} (-3H\dot{\phi}^2) = -3H\dot{\phi}^{2} \rho_{\phi}
\end{equation}
\vspace{2mm}

\noindent
By using the definition of the Hubble parameter $H = {\dot{a}\over{a}}$, one can write
\begin{equation}
3H^{2} = -{1\over{a}}{da\over{d\phi}}{d\rho_{\phi}\over{d\phi}}{1\over{\rho_{\phi}}}
\end{equation}
\vspace{2mm}

\noindent
Substituting the above equation in the Hubble equation (1), one can write
\begin{equation}
 {da\over{d\phi}}{d\rho\over{d\phi}} = -\kappa^{2} a \rho^{2} L^{2}(\phi),
\end{equation}
\vspace{2mm}

\noindent
where $\rho$ is the total energy density of the universe and can be written as $\rho = {\rho_{\phi}\over{\Omega_{\phi c}}}$. Defining a new variable $b(\phi)$ as
\begin{equation}
b(\phi) = exp\left[\int^{\rho} d\rho {1\over{\rho^{2}L^{2}(\rho)}}\right],
\end{equation}
\vspace{2mm}

\noindent
one can write equation (18) as
\begin{equation}
{da\over{d\phi}}{db\over{d\phi}} = - \kappa^{2} ab,
\end{equation}
\vspace{2mm}

\noindent
which can be subsequently integrated to give the scale factor $a$:
\begin{equation}
a = Exp\left[ \int -\kappa^{2} ({db\over{d\phi}})^{-1} b d\phi \right].
\end{equation}
\vspace{2mm}

\noindent
Using the equation (16), one can also calculate the time dependence of the scale factor through the following equation:
\begin{equation}
t = -\sqrt{3}\kappa \int d\phi \rho^{3/2}(\phi) L(\phi) ({d\rho\over{d\phi}})^{-1}
\end{equation}
\vspace{5mm}

\noindent
\section{Different Classes of Modified Gravity}
\subsection{Randall-Sundrum model}
For Type-II Randall Sundrum brane-world model where 3 brane with positive tension is embedded in five-dimensional Anti de-Sitter spacetime, the modification to the standard gravity is given by
\begin{equation}
L(\rho) = \sqrt{1 + {\rho\over{2\sigma}}},
\end{equation}
\vspace{2mm}

\noindent
$\sigma$ is the tension of the 3-brane. One can now use equations (5) and (15) to determine the scaling potential which  is given by:
\begin{equation}
V(\phi) = {4\sigma Y_{c}^{2}\over{\sigma\lambda_{c}^{2}Y_{c}^{2}\kappa^{2}\phi^{2}-2}}
\end{equation}
\vspace{2mm}

\noindent
One can also obtain the time dependence of the scale factor $a(t)$ using equations (21) and (22):
\begin{equation}
a(t) = \left({\kappa^{2}\lambda_{c}^{4}Y_{c}^{4}\sigma\over{6}}t^{2} - 1\right)^{1\over{\lambda_{c}^{2}Y_{c}^{2}}}.
\end{equation}

As explained in section II, there are two stable solutions. One of them can not produce a viable late time dark energu model, as the background fluid can never behave like matter. For the other case, one can have a late time scalar field dominated case ($\Omega_{\phi}=1$) with accelerated expansion. We have plotted the time evolution for this case in figure 1. We have taken ${\sigma\over{m_{p}^2}}= 10^{-20}$.We have shown the evolution for different values of $\lambda$. We should mention that evolution of the universe is same as obtained by Copeland {\it et al.} in \cite{copeland2} for a standard canonical scalar field. But the due to the noncaninical nature of the kinetic term, the required potential is different. In this case the acceleration occurs for $\gamma_{\phi} <2/3$ in the late time when the $\rho$ term in the action dominates whereas for $\gamma_{\phi} < 1/3$, acceleration occurs in early time when the $\rho^2$ term dominates. This translates to the constraints on $\lambda_{c}$ as $\lambda_{c} < 1.86$ and $\lambda_{c} < 1.01$ respectively.
\begin{figure}[t]
\centerline{\epsfxsize=3.7truein\epsfbox{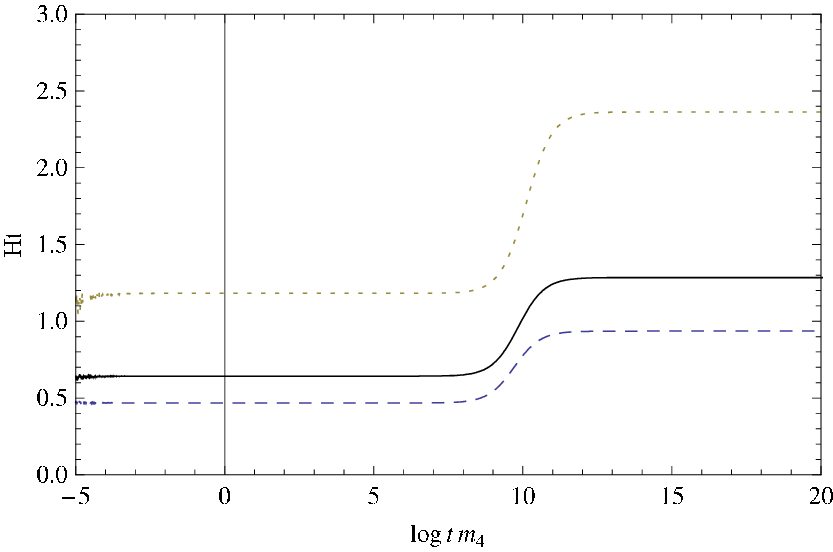}}
\caption{The evolution of the universe for Randall-Sundrum II brane world models. We have taken
${\sigma\over{m_{p}^4}} = 10^{-20}$ for this figure. $\lambda_{c} = 1, 1.5,2$ from top to bottom.}
\end{figure}

\vspace{5mm}

\subsection{Shtanov-Sahni Braneworld Cosmology}
There is another interesting class of models proposed by Shtanov and Sahni where a 3 brane with negative tension $\sigma$ is embedded in a five dimensional conformally flat space, where the fifth dimension has time-like signature. In this case the modification to standard gravity is describe as:
\begin{equation}
L(\rho) = \sqrt{1- {\rho\over{2|\sigma|}}}
\end{equation}
\vspace{2mm}

\noindent
One can calculate the potential which is given by 
\begin{equation}
V(\phi) = {Y_{c}^{2}\over{{1\over{2|\sigma|}} + {\kappa^{2}\lambda_{c}^{2}Y_{c}^{2}\over{4}}\phi^{2}}}.
\end{equation}
\vspace{2mm}

\noindent
One can also calculate the corresponding scalar factor $a(t)$
\begin{equation}
a(t) = \left({\kappa^{2}\lambda_{c}^{4}Y_{c}^{4}|\sigma|t^{2}\over{6}} + 1\right)^{1\over{\lambda_{c}^{2}Y_{c}^{2}}}
\end{equation}
\vspace{5mm}

\subsection{Cardassian model}
All the above modified gravity models modifies the gravity at small distance scales i.e it modifies gravity at high energy scales (early times).  But recently SnIa, CMB and LSS measurements have confirmed the late time accelerated expansion of the universe. A number of phenomenological models have been proposed to explain this late time accelerated expansion through modifications of standard gravity at large distance scales. In one such model, know as ``Cardassian model'' originally proposed by Freese and Lewis the modification is given by
\begin{equation}
L(\rho) = \sqrt{1+A\rho^{n}}
\end{equation}
\vspace{2mm}

\noindent
The present acceleration of the universe can be obtained in such model with $n < -1/3$ when the universe contains only the matter. Although the presence of any scalar field was not assumed in this original model, it is interesting to study the scaling behavior of the background cosmology in the presence of both matter and a scalar field, a noncanonical one for our present purpose. To get the result analytically in closed form we have assumed $n= -1/2$ in the following calculations. We should mention that taking different values of $n$ is always possible, only problem is that solution may not be in a closed form. We have used this particular value of $n$ to present the solution in a closed form. With this choice of $n$, one can directly obtain the form of the potential $V(\phi)$ and the corresponding scale factor $a(t)$ as following:
\begin{equation}
V(\phi)  = {A^{2}Y_{c}^{2}\over{\left[{\lambda_{c}^{2}\kappa^{2}Y_{c}^{2}A^{2}\phi^{2}\over{16}}-1\right]^{2}}}
\end{equation}
\vspace{5mm}
\begin{equation}
a(t) = \left[A\left({Y_{c}^{2}\lambda_{c}^{2}\kappa \over{4\sqrt{3}}}\right)^{2}t^{2}-1\right]^{2\over{Y_{c}^{2}\lambda_{c}^{2}}}.
\end{equation}

We have shown the evolution of universe for this in figure 2. We have assumed $A/m_{4} = 10 ^{-10}$. Acceleration of the universe takes place in early time  for $\gamma < {2/3}$, which constrains the $\lambda_{c}$ parameter as $\lambda_{c} < 1.86$. For late time, universe always accelerates.

\begin{figure}[t]
\centerline{\epsfxsize=3.7truein\epsfbox{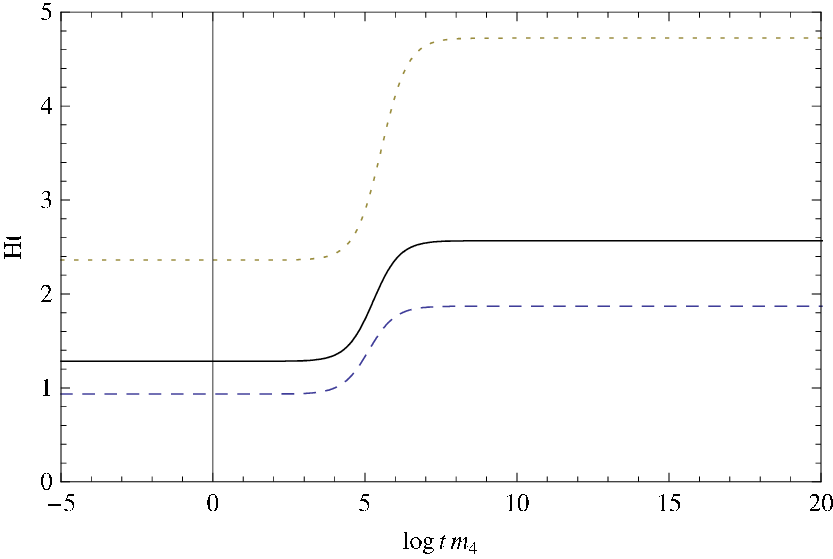}}
\caption{The evolution of the universe for Cardassian model. We have taken
${A\over{m_{p}^4}} = 10^{-10}$ for this figure. $\lambda_{c} = 1, 1.5,2$ from top to bottom.}
\end{figure}

\subsection{Dvali-Gabadadze-Porrati Brane-World Gravity Model}
The Dvali-Gabadadze-Porrati (DGP) model is a brane-world model (a 3-brane embedded in 5D anti-deSitter bulk) with the incorporation of the scalar curvature term for the 3-brane in the total action. The bulk is empty and all kind of matter fields are restricted to stay only on the 3-brane. The modified FRW equation for DGP model is given by
\begin{equation}
H^{2} \pm {H\over{r_{0}}} = {8\pi G\over{3}}\rho
\end{equation}
\vspace{2mm}

\noindent
where $r_{0}^2 = {m_{4}^{2}\over{2 m_{5}^{3}}}$ with $m_{4}$ and $m_{5}$ being the 4-d and 5-d Planck mass respectively. The parameter $r_{0}$ actually determines the scale at which the switchover from standard gravity to modified one takes place.The (+) and (-) sign corresponds to different kinds of embedding the 3-brane in the 5d bulk. The modified equation now becomes
\begin{equation}
H =   {1\over{2r_{0}}} \left[\mp 1 +\sqrt{1+\alpha_{1} \rho}\right]
\end{equation}
\vspace{2mm}

\noindent
where $\alpha_{1} = {32 \pi r_{0}^{2}\over{3m_{4}^{2}}}$. A direct comparison of the above equation with equation (1), results the correction term in the Friedmann equation as
\begin{equation}
L = {1\over{\sqrt{\alpha_{1}\rho}}}\left[\mp 1 +\sqrt{1+\alpha_{1}\rho}\right].
\end{equation}
\vspace{2mm}

\noindent
Putting this in the equation (15), one can integrate to get
\begin{equation}
{\sqrt{\alpha_{1}}\over{\mp 1+\sqrt{1+\alpha_{1}\rho}}} + \sqrt{\alpha_{1}}Sinh^{-1}\left(1\over{\sqrt{\alpha_{1}\rho}}\right) = \kappa\lambda_{c}Y_{c}\phi
\end{equation}
\vspace{2mm}

\noindent
This equation together with equation (5) can be used to find the scaling potential $V(\phi)$. One can also calculated the corresponding time dependence. This is given by
\begin{equation}
{3X_{c}^{2}\over{2 r_{0}}} t = Cotanh^{-1}\sqrt{1+\alpha_{1}\rho} + {1\over{\sqrt{1+\alpha_{1}\rho}\mp 1}}
\end{equation}
\vspace{5mm}

\section{Conclusion}
The issue of late time acceleration of the universe has been one of the most serious challenges in cosmology nowadays. While including an extra dark energy component with repulsive gravity in the energy budget of the universe, is one of the most studied approaches for explaining, modifying Einstein's gravity at large distance scales, has also been taken seriously in recent times. While any modifications to Einstein's gravity has its own problems, this idea has been rigorously pursued by various researchers in recent times. One of the main motivations of such idea is that, modifications of Einstein's gravity can arise at different energy scales very naturally through compactifying higher dimensional theories. 

On the other hand, the concept of scaling solutions in cosmology has been taken seriously in recent times, as such solutions are necessary for solving cosmic coincidence problem in dark energy models.  In this paper, we have studied in systematic way the scaling solutions in modified gravity models when the universe contains a tachyon type scalar field in addition to the sandard matter field. This is an extension of the earlier work done by Copeland et al.\cite{copeland2} where the scalar field with a canonical scalar field was considered.

We first describe the general equations and method of calculating scaling potential. Later on, we take specific modified gravity models and apply our method. We take four choices of modifications, e.g the Randall-Sundrum II model, Shtanov-Sahni Model, Cardassian model and DGP model. Our method can also be used to consider any modified gravity to calculate the scaling potential, while considering the tachyon type scalar field. We should mention that Tsujikawa and Sami\cite{ss} have earlier considered scaling solutions in modified gravity with a tachyon field. But for the modification, they have considered the special case when $H^{2} \propto \rho^{n}$ which does not include modifications like DGP model. Das et al.\cite{rupam} have also considered tracking solutions in modified gravity both with quintessence and K-essence type fields. Their approach is different from ours and they have not considered fields like tachyon which has some specific features. 

In future, one can take the general K-essence action for the non-canonical scalar field, and calculate the scaling potential. This will be our future goal. 
  
\section{Acknowledgement}
The authors acknowledge the financial support provided by the University Grants Commission, Govt. Of India, through the major research project grant (Grant No:  33-28/2007(SR)).

\end{document}